# Verification of low-frequency signal injection method for earth-fault detection

Nina Stipetic, Bozidar Filipovic-Grcic, Igor Ziger, Silvio Jancin, Bruno Jurisic, Dalibor Filipovic-Grcic, Alain Xémard

*Abstract*--**Unearthed neutral is commonly used in networks which require continuous power supply. This is common in MV circuits of industrial and power plants. Unearthed networks can remain in operation during an earth-fault, but fast determination of the faulty line is key for prevention of further fault escalation. Signal injection is one of the fault location methods often used in LV unearthed networks. The possibility of applying this method in MV networks depends on how to inject the signal into unearthed phases. In such networks, it is possible to use a group of three inductive voltage transformers (IVTs) for signal injection. After the simulations have shown promising results of signal injection and earth-fault detection in MV network, an experimental test was performed. This paper describes the experimental setup and shows the measurement results of signal injection method at MV level supported by EMT simulations.**

## I. Introduction

THE grounding method is important for the reliability of the power system networks' operation. Depending on the grounding method, different waveforms and amplitudes of overvoltages and fault currents may appear during earth-faults. In unearthed networks, the earth-fault current closes through systems' capacitances to ground which does not lead to a high fault current and there is no need for fault interruption. That is why an unearthed neutral is often used in industrial networks, and in networks within power plants (i.e. nuclear power plants) which require continuous power supply. In such networks, earth-faults are the most frequent fault type, representing 50-90 % of all faults [1]-[3]. Even though an unearthed network can remain in operation during an earth-fault, fast determination of the faulty line is the key for prevention of further fault escalation.

Extensive research and development efforts are continuously focused on new methods for earth-fault location in distribution networks. Most of these methods rely on either centralized or decentralized measurements of fundamental-frequency current and voltage, as well as impedance and fault distance calculations [4], [5]-[8].

Research has also explored the traveling-wave-based method, which theoretically shows promising results. However, its practical application is hindered by the complexity of MV networks and the need for advanced measurement equipment with high sampling rates. Both transient and high-frequency measurement techniques encounter similar challenges in measurement [9]-[14]. Artificial intelligence-based methods have recently gained significant attention, but examples of their real-world application remain scarce [15]-[17].

Signal injection is one of the fault location methods often used in LV unearthed networks. The possibility of application of this method in MV networks depends on how to inject the signal into unearthed phases with voltages ranging from 5 to 35 kV. In isolated MV networks there is usually a group of three single- phase insulated inductive voltage transformers (IVTs), whose tertiary delta circuit is closed over a resistor for ferroresonance suppression. This group of three IVTs can be used for the signal injection. Simulations on an example of industrial, radial cable network and mixed cable-overhead line distribution network in [18]-[19] have shown that injection is possible through IVTs. The injected signal is traceable in the residual currents in the network. Since it closes its path via fault location, its content in the residual current on the faulty feeder is expected to be the highest. It was also shown that in some cases it is possible to use existing, standard IVTs for the injection, but it is also possible to influence the core and primary winding design to make the IVT more suitable for both measurement and signal injection.

### A. Working principle of signal injection method

Fig. 1 shows the signal injection method operating principle.

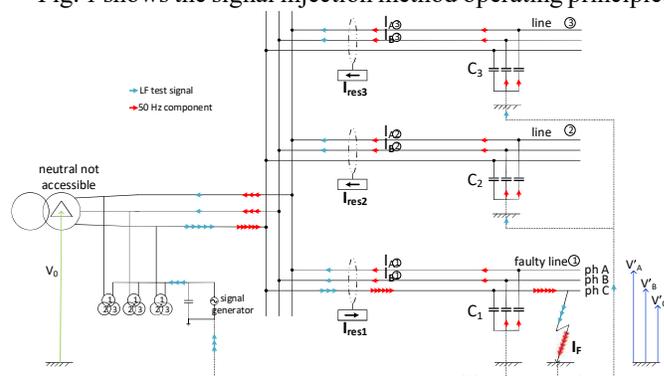

Fig. 1. Signal injection working principle

The signal generator is connected between the system's artificial neutral (at the star connection of primary windings of IVTs) and ground. When an earth fault is indicated by basic protection functions (i.e. zero-sequence voltage occurrence),

N. Stipetic and B. Filipovic-Grcic are with University of Zagreb Faculty of Electrical Engineering and Computing, 10000 Zagreb, Croatia (e-mail: nina.stipetic@fer.hr, bozidar.filipovic-grcic@fer.hr).
I. Ziger is with Koncar – Instrument Transformers Inc., Josipa Mokrovica 10, 10090 Zagreb (e-mail: igor.ziger@koncar-mjt.hr).
S. Jancin, B. Jurisic, D. Filipovic-Grcic are with KONCAR – Electrical Engineering Institute, 10000 Zagreb, Croatia (e-mail: sjancin@koncar-institut.hr, bjurisic@koncar-institut.hr, dfilipovic@koncar-institut.hr).
A. Xémard is with the Department of Research and Development, Électricité de France, 91120 Palaiseau, France (e-mail: alain.xemard@edf.fr).
Paper submitted to the International Conference on Power Systems Transients (IPST2025) in Guadalajara, Mexico, June 8-12, 2025.

the low-frequency signal is injected into the system. It is also possible to plan periodical injections and observe the changes in the LF signal measurements throughout the network. The injected signal closes its path via fault location. The capacitances of healthy feeders represent a high impedance path for low frequency test signal. Hence, the probability that the injected signal will close through fault location is higher if its frequency is lower.

### B. Signal injection circuit

Fig. 2 shows the more detailed scheme of the signal injection circuit.

The secondaries of three-winding IVTs are used for measurement and are loaded with rated load. The tertiary is delta-connected and loaded with resistor for dampening the ferroresonance phenomena [20]. A signal generator injects current through primary windings of IVTs which are grounded over a capacitor to avoid current being shunted. A similar circuit is used for Permanent Insulation Monitors (PIMs) connection, which uses DC injection for calculation of total network resistance. For PIM connection, the advisable value for the capacitor is 2.5 µF with an insulation voltage of 1600 V [1].

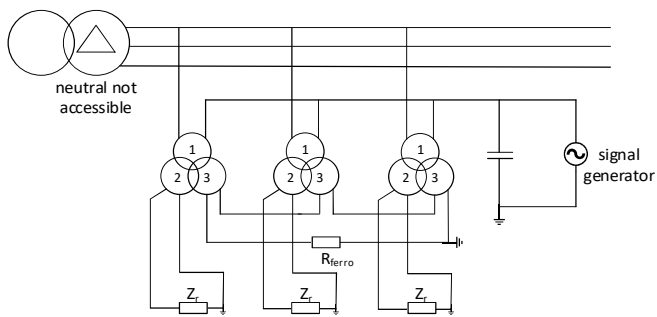

Fig. 2. Schematic preview of signal injection into isolated MV network using the primary windings of IVTs

Induced voltage in IVT is described by the following equation:

$$E_{rms} = 4.44 \cdot N \cdot f \cdot \Phi_{max}; \quad \Phi_{max} = \frac{E_{rms}}{4.44 \cdot f \cdot N} \quad (1)$$

where $\phi$ is the core magnetic flux, $E$ is the induced voltage and $N$ is the number of turns. Observing the equation (1) it is clear that changing the ratio $E/f$ leads to a change in the flux $\phi$. Increase of the flux over the knee point of the current-flux curve will lead to core saturation. Saturation increases the current through the primary winding and distorts the measurement on secondaries. Long-duration saturations should generally be avoided, but short saturations are acceptable as long as the primary winding is not thermally endangered. Hence, the IVT is the key piece of equipment for the application of signal injection method in unearthed networks. Signal parameters should be chosen optimally for injection into the network. The practice at LV is to use a sinusoidal signal with frequency of 2.5 Hz. Lower frequencies are more favourable in terms of LF signal distribution towards the fault location, since the line-to-ground capacitive reactances and high insulation resistances of healthy feeders represent highly resistive path for the injected signal. A DC injection would be the best in this regard, however, in that case a DC current sensor or DC component calculation in residual currents is needed. Additionally, the DC injection is most unfavourable in terms of IVT saturation. Nevertheless, in [19] it was shown that IVT construction can be optimized to avoid saturation related to the earth-fault and signal injection. In the same manner, the resistor for avoiding ferroresonance can be adjusted if necessary, which can be checked by simulations. The risk of ferroeresonance only increases during the saturation periods, which can be minimised by IVT design.

## II. LABORATORY EXPERIMENTAL PROOF OF SIGNAL INJECTION METHOD THROUGH IVTs

### A. Experimental setup description

For the experimental proof of the described signal injection through IVTs, three standard MV IVTs, type 4VPA1-12, were used for laboratory experimental test. Fig. 3 shows the schematic of the test circuit, and Fig. 4 the photo of the laboratory test arrangement, with designations and explanations for the main equipment used.

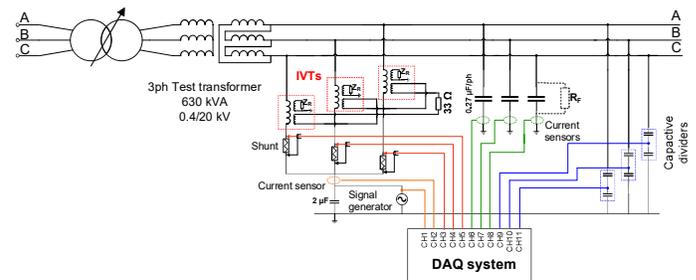

Fig. 3. Schematic of the laboratory test setup

The three-phase experimental setup consists of three-phase regulating and test transformer used for increasing the voltage to MV level of 10 kV. Capacitors of 0.27 µF are used to represent the network to ground capacitance, which is approximately a concentrated equivalent of 1 km long cable. An earth-fault is intentionally made by shorting the capacitor in one of the phases to ground. The signal-injection circuit is connected to phases through the IVTs, following the layout given in Fig. 2. The secondaries were connected to the rated burden with cos$\varphi$=0.8 and the 33 Ω resistor was connected in the tertiary delta circuit.

During the experimental earth-fault and signal injection, the phase voltages were measured using the capacitive dividers, the primary winding currents were measured using shunts, the currents through phases and capacitors are measured by flux-gate current sensors. The residual current was calculated mathematically based on measured phase currents. Additionally, the injection voltage and current were measured for control, as well as the current in the tertiary delta winding. Fig. 5 shows the measured $U$-$I$ curves of used IVTs and the calculated $\phi$-$I$ curves. Despite the same nameplate data and type the magnetization curves reveal slight difference of the curve for the IVT placed in phase B, while the curves of IVTs in phase A and C overlap very well.

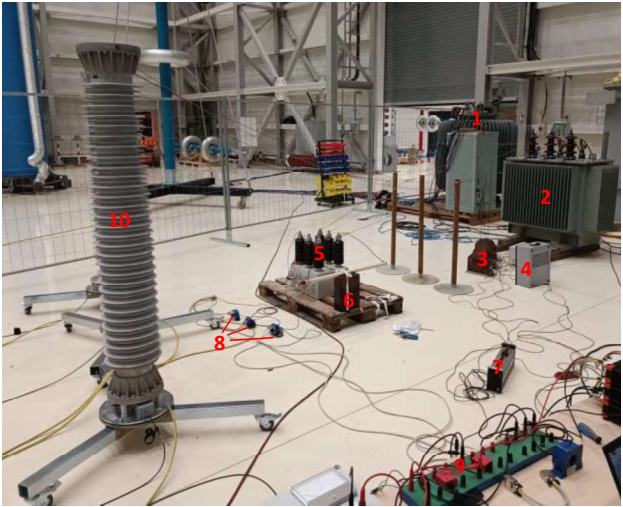

| 1 | 3ph Regulating transformer, 500 kVA, 0-420 V |
|---|---|
| 2 | 3ph Test transformer, 630 kVA, 0.4/20 kV |
| 3 | IVTs |
| 4 | Rated burden on IVT secondary |
| 5 | MV capacitors – cable capacitance equivalents |
| 6 | Earth-fault spot where fault resistances are connected |
| 7 | Resistor for ferroresonance damping in tertiary circuit |
| 8 | Measurement of currents through MV capacitors |
| 9 | Primaries star connection and the LV capacitor |
| 10 | Capacitive dividers for voltage measurement |

Fig. 4. Photo of the laboratory test arrangement

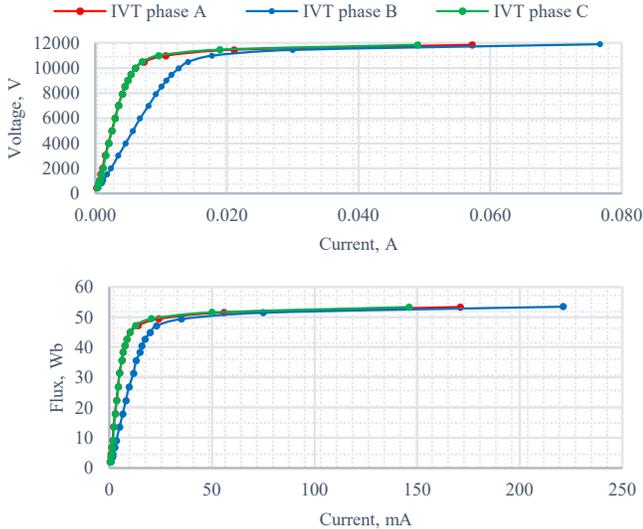

Fig. 5. Measured *U-I* curves and calculated *ϕ-I* curves of the three IVTs used in the experiment

### III. MEASUREMENT RESULTS

To avoid switching operation and possibility of triggering ferroresonance phenomena, an earth-fault in the circuit was connected off-line, prior to rising the voltage. Firstly, only earth-fault was considered for measurement to check if all conditions are as expected. Afterwards, the injection circuit was connected to the setup. Both AC and DC injections were tested. The AC injection was limited by the equipment and the highest injected amplitude was 100 $V_{peak}$. During the AC injection, the injected signal frequency was changed. During the DC injection, the limitation was the thermal stress of the primary windings of IVTs. Both AC and DC injections were done for a solid earth-fault and for an earth-fault with several different fault resistances. In this paper, the results for solid earth-fault and for the case of 690 Ω at fault location are presented.

### A. Measurements during a solid earth-fault

First measurements were taken without the signal injection, to check if the voltages and currents during an earth-fault behaved as expected. The measured waveforms are shown in Fig. 6-Fig. 9. The voltage was raised to 10.5 $kV_{rms}$ which corresponds to the line voltage value during normal, no-fault condition. As expected, the measured phase-to-ground voltages in healthy phases are equal to line voltage value of 10.5 $kV_{rms}$. In phase C, where the earth-fault occurs, the voltage dropped to zero. The primary winding currents of IVTs reach values of 35 $mA_{rms}$, 30 $mA_{rms}$ and 16 $mA_{rms}$ in phases A, B and C respectively. Since the earth-fault introduces asymmetry, there is current in tertiary winding, and it equals 3.1 $A_{rms}$. The total earth-fault current in any unearthed network is calculated according to:

$$I_f = 3\omega C_{ph} V_{ph} \quad (2)$$

where $C_{ph}$ is the capacitance-to-ground per phase and $V_{ph}$ is the phase voltage value prior to the fault [1]. In this case, the fault current equals 1.654 A.

The initial test confirmed that the measurements are as expected, and that the experimental setup is correct. The next measurement included the signal generator for signal injection.

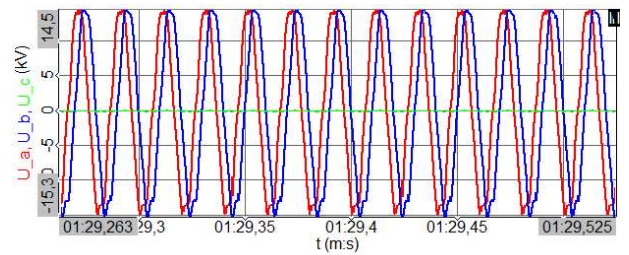

Fig. 6. Recorded voltages during a solid earth-fault, in healthy phases voltage reaches line value, in phase C where the fault is applied voltage drops to zero

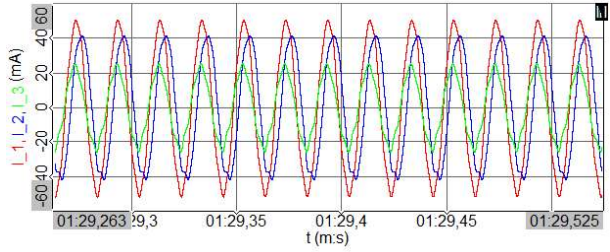

Fig. 7.  Recorded primary winding currents during a solid earth fault (35 mA$_{rms}$, 30 mA$_{rms}$ and 16 mA$_{rms}$)

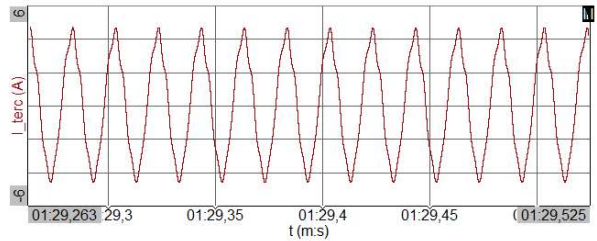

Fig. 8.  Recorded tertiary winding current during a solid earth fault (3.1 A$_{rms}$)

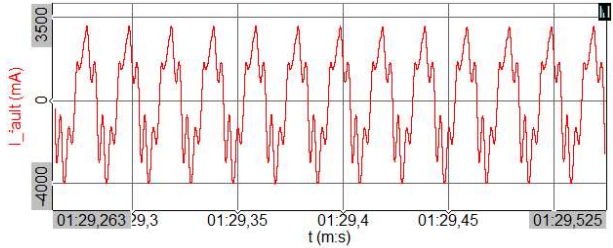

Fig. 9.  Recorded fault current during a solid earth-fault, $I_{rms}$ = 1.6 A

*B. Measurements during a solid earth-fault and LF sinusoidal signal injection*

After confirmation of earth-fault conditions, the signal generator was included in the circuit. Fig. 10 shows the waveform of primary currents during the simultaneous earth-fault and injection of signal 2.5 Hz, 100 V$_{peak}$. Fig. 11 shows the residual current (*3I$_0$*). In these waveforms, both 50 Hz and 2.5 Hz are present, and according to Fourier transform (FFT) analysis, the 2.5 Hz component in the residual current equals 19 mA and the 50 Hz component 105 mA.

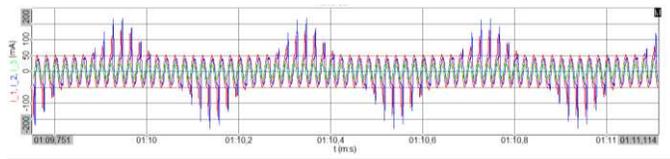

Fig. 10.  Recorded primary winding currents during a solid earth fault and injection of the signal 2.5 Hz, 100 V$_{peak}$

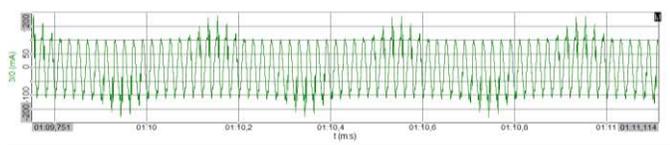

Fig. 11.  Recorded residual current during a solid earth fault and injection of the signal 2.5 Hz, 100 V$_{peak}$

For the same amplitude of the injected signal, lowering the frequency means increasing the flux in the core, thus higher saturation is expected. At the same time, a higher amplitude of the LF component is expected in the residual current. Fig. 12 and Fig. 13 show the primary winding currents and the residual current during the injection of 0.5 Hz, 100 V$_{peak}$ signal. In this case, according to FFT analysis, the 0.5 Hz component in the residual current equals 89 mA. The general influence of increasing the frequency of injected signal on the waveform of the residual current is shown in Fig. 14 - Fig. 15. For the injected signal with variations in the frequency (Fig. 14), the change in the residual current is shown in Fig. 15. FFT analysis for this case of varying the signal injection frequency is given in Fig. 16. It depicts the drop of the LF component in the residual current upon the increase of frequency of the injected signal.

These measurements have confirmed the possibility of LF signal injection through the IVTs and the traceability of the injected signal in the frequency content of the residual current. They have also confirmed the principle of how the change in injected signal frequency affects the primary currents of IVTs and low-frequency content in residual current.

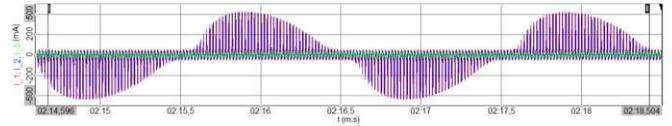

Fig. 12.  Recorded primary currents during the solid earth fault and injection of the signal 0.5 Hz, 100 V$_{peak}$

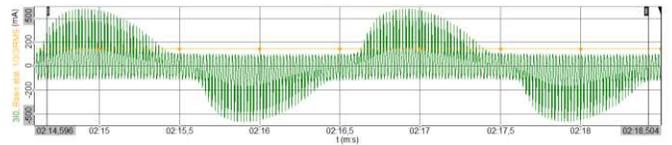

Fig. 13.  Recorded residual current during a solid earth fault and injection of the signal 0.5 Hz, 100 V$_{peak}$

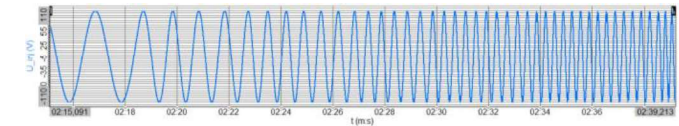

Fig. 14.  Recorded injected signal with amplitude of 100 V$_{peak}$ and frequency ranging from 0.5 Hz to 3 Hz

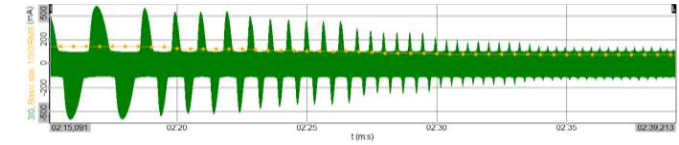

Fig. 15.  Recorded waveform of residual current for the injection of the LF signal from Fig. 14

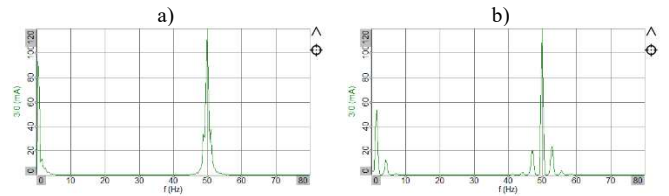

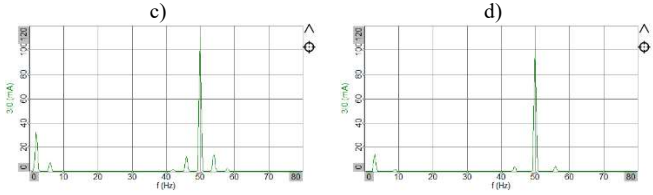

Fig. 16. FFT during the injection of signal from Fig. 14. a) 0.4 Hz, 93 mA, b) 1.5 Hz, 53 mA, c) 2 Hz, 32 mA, d) 3 Hz, 14 mA

### C. Measurement during a solid earth-fault and DC signal injection

After the LF sinusoidal injection, the measurements were taken during the DC signal injection. Theoretically, the DC injection is the most favorable in terms of the probability of the signal closing through the fault location. However, it will cause the highest saturation and the greatest distortion of the primary winding currents, and it requires special DC sensors or calculation of mean value of residual currents throughout network. In this experiment, the DC signal was increased up to 225 V. During the 225 V injection, the primary winding currents reached 275 mA$_{rms}$, 267 mA$_{rms}$ and 121 mA$_{rms}$, in phases A, B and C respectively. These current values do not represent a threat to the primary windings if the stress is short-term. For longer DC injections, the IVT's core and primary winding could be redesigned as discussed in [19]. Fig. 17 shows the recorded injected DC signal and Fig. 18 the residual current during this DC injection.

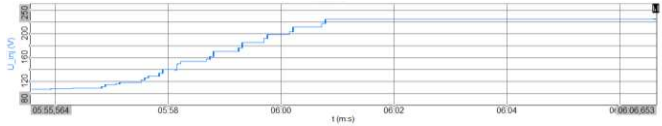

Fig. 17. Recorded injected DC signal, variation of values from 100 V to 225 V

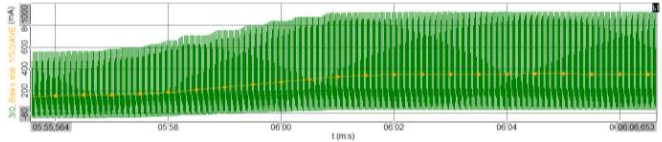

Fig. 18. Recorded waveform of residual current for the injection of the DC signal from Fig. 17

The injected signal in the residual current is traceable as its DC component, and it rises from 160 mA for the 100 V injection to 357 mA for 225 V injection. Compared to LF sinusoidal injection, the DC component in residual current is higher than the LF content in the residual current in case of AC sinusoidal injection. Hence the DC injection is more effective. However, for some networks, the solution will depend on the fact if the existing IVT are to be used for injection or new ones that could be adapted for signal injection, and the signal detection possibility.

### D. Measurements during an earth fault with fault resistance of 690 Ω

After the solid earth-fault, an earth-fault with fault resistance was considered. For this purpose, the capacitor in phase C was shorted over a 690 Ω resistor. Fig. 19-21 show the measured waveforms of primary winding currents, the residual current and the voltages recorded in this case. Comparing them to the solid earth-fault case, the waveforms are similar. However, it can be observed that in case with the fault resistance, only IVT in phase B gets saturated. The resistance at the fault location decreases the zero-sequence voltage and changes the phase voltage values. Consequently, the total fault current is decreased and the saturation in different phases is dictated by the new phase voltages conditions. According to the FFT analysis, the 2.5 Hz component in residual current is 18.8 mA, while the 50 Hz component equals 98 mA, which is lower than in the solid earth-fault case. Fig. 21 shows that, due to existence of the resistance at the fault location in phase C, the voltage in phase C is not zero, and the voltage in phase B is slightly higher than in phase A, which causes saturations of IVT in phase B only. The slight difference in magnetization curve of the IVT in phase B that is shown in Fig. 5 also contributes to this difference in saturation.

The DC injection was also repeated in case with the fault resistance. In this case, the amplitude of the DC signal could be increased to higher values than in case of a solid earth-fault, due to the existence of the resistor and dampening the primary currents and the saturation. The injection was done up to 330 V. For the injection of 200 V the DC component in the residual current was 141 mA, and for the 330 V it raised to 245 mA. During the 330 V injection, the primary winding currents of the IVTs were 199.5 mA$_{rms}$, 195.1 mA$_{rms}$ and 107.9 mA$_{rms}$.

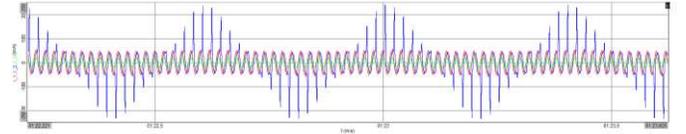

Fig. 19. Recorded waveform of primary winding currents during an earth-fault with 690 Ω fault resistance and signal injection 2.5 Hz, 100 V$_{peak}$

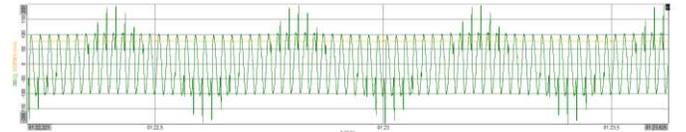

Fig. 20. Recorded waveform of residual current during an earth-fault with 690 Ω fault resistance and signal injection 2.5 Hz, 100 V$_{peak}$

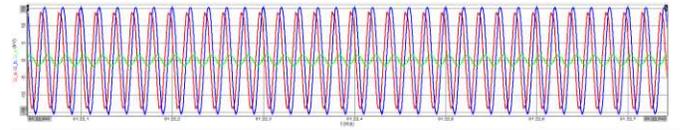

Fig. 21. Recorded waveform of voltages in case of an earth-fault with 690 Ω fault resistance and 2.5 Hz, 100 V$_{peak}$ injection

Again, the possibility of LF AC and DC signal injection and its traceability in the residual current was confirmed in the case with fault resistance existence.

### E. Comparison with simulation results

For analyzing the possibility of signal injection and earth-fault detection in real networks, EMT simulations should be performed as demonstrated in [18], [19]. In this paper, laboratory setup was modelled in EMT to check the accuracy of the model by comparing simulation results with measurements. Thus, the same setup from Fig. 3 was modelled in EMT. The measured voltages and the injected signal were extracted from the measurements and imported to the simulation model. The comparison of the measurements and simulation results for the case of 2.5 Hz, 100 V<sub>peak</sub> injection in case with fault resistance of 690 Ω is given in Fig. 22 - Fig. 25.

The measured and simulated waveforms overlap perfectly when there is no saturation. However, when highly saturated, the simulated results lead to higher peak values of primary winding currents and consequently the peaks of residual current. There is also an asymmetry noticed in positive and negative peaks. The primary windings currents are highly sensitive to the magnetization curve model and the distortion of the system voltage, presence of harmonics and their phase angles as well as the phase difference of the injected signal to the voltage in the network are all the factors that influenced these differences. Nevertheless, it can be concluded that all waveforms are in good accordance. As mentioned before, the aim should be to avoid the saturation in any case, which is possible even during the signal injection if the IVT design is slightly changed, as suggested in [19]. After this experimental verification, the optimization of the IVT design for the signal injection is a subject for further research.

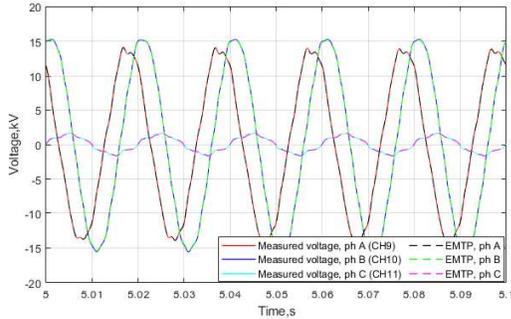

Fig. 22. Comparison between EMT simulations and phase voltages measured over capacitor dividers, corresponding to CH 9, CH 10 and CH 11 from Fig. 3

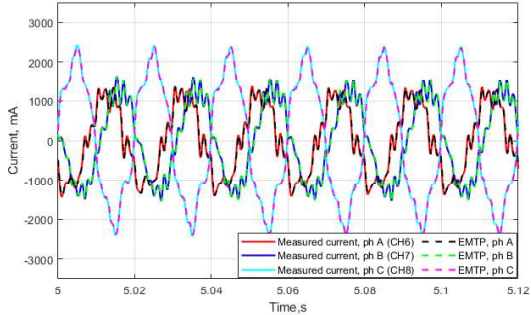

Fig. 24. Comparison between EMT simulations and measurements of currents through capacitors, corresponding to CH 6, CH 7 and CH 8 from Fig. 3

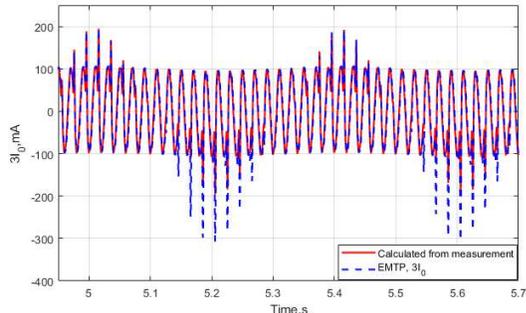

Fig. 25. Comparison between EMT simulations and measured residual current, corresponding to sum CH 6 + CH 7 + CH 8 from Fig. 3

## IV. CONCLUSIONS

This paper shows the results of the experimental test of LF and DC signal injection in the MV isolated network, using the primary windings of inductive voltage transformers. The signal injection can be used to detect the earth-fault and the faulty line in unearthed network, by tracing the injected signal in residual currents throughout the network.

The experiment was conducted with available IVTs of the same type and the same nameplate data. The test setup consists of one isolated line with connected capacitors to ground which represent the cable capacitance which plays a key role in zero-sequence circuit when an earth-fault occurs. An earth-fault was simulated by shorting one of the capacitors. During the experiment, the voltages, the primary winding currents, the injected voltage and the currents through capacitors were measured. The residual current was calculated based on measured currents through capacitors. It was shown that it is possible to inject the LF and DC signals through the primary windings of IVTs and that the injected signal is traceable in the residual current. The influence of changing the injected signal frequency was shown. The recorded measurements in case of a solid earth-fault and an earth fault over a 690 Ω resistor are presented.

In case of a solid earth-fault and 2.5 Hz, 100 V injection, the 2.5 Hz component in the residual current is the highest. Increasing the amplitude of the injected signal would increase the 2.5 Hz component in the residual current. The frequency of the injected signal was changed. It was shown that the LF component in the residual current drops as the frequency of the injected signal increases. Increase of the frequency lowers the LF component in the residual current and makes it harder to detect. DC signal injection was also tested. Compared to the LF signal injection, the DC injection is more favourable in terms of signal detection, but care should be taken regarding the primary winding thermal stress. In case of a fault over a resistance, the experiment showed similar results.

For analysing the possibility of signal injection and earth-fault detection in real networks, EMT simulations should be performed. The key importance for simulations is that EMT model has to consider all ground capacitances in the network as well as all possible ground paths for the injected signal (other IVTs in the network, star connected loads), since these are the factors that affect the method the most and may affect the possibility of its application. In this paper, laboratory setup was modelled in EMT to check the accuracy of the model by comparing simulation results with measurements. Comparison proves that any future simulations related to the injection method should yield accurate results.

This experimental test has proven that the injection of both LF AC signal and DC signal in the unearthed network is possible using the group of three standard 10 kV IVTs. The results show that the DC injection is more favourable in terms of tracing the injected signal in the residual current. The impact of DC injection on the IVTs is important and care should be taken not to thermally overload the primary windings. If injection of such signal that causes excessive saturation is needed for faulty feeder detection, re-design of the IVT's core should be considered in order to adjust the IVT for such injections. Future research includes the validation of all the measurement results by simulations and optimizing the IVT for DC injection as suggested in [19].

In some HVDC systems, especially in monopolar configurations, the neutral point is kept ungrounded or isolated from the earth. This prevents fault currents from flowing to the ground, but special protection mechanisms are needed to deal with fault conditions, such as insulation failures. In some cases, an ungrounded system can lead to more complex fault detection and higher risk of damage to the equipment. The method presented in this paper might find its applications also for fault detection in DC networks, but this topic will be investigated in future work.